%% file: Proceedings.tex
\documentclass[a4paper]{jpconf}
\usepackage{graphicx}

\usepackage{amsmath,amsfonts,amssymb,amsthm}
\usepackage{mathrsfs}
\usepackage{graphicx}
\usepackage{bbm}
\usepackage[small]{caption}
\usepackage[utf8]{inputenc}
\usepackage[small,loose]{subfigure}  
\usepackage{pifont} 

\newcommand{\maN}{\ensuremath{\mathcal{N}}}
\newcommand{\maT}{\ensuremath{\mathbb{T}}}

\newcommand{\maG}{\ensuremath{\mathcal{G}}}
\newcommand{\rep}[1]{\ensuremath{\boldsymbol{#1}}}
\newcommand{\Z}[1]{\ensuremath{\mathbb{Z}_{#1}}} 
\newcommand{\E}[1]{\ensuremath{\mathrm{E}_{#1}}} 
\newcommand{\SU}[1]{\ensuremath{\mathrm{SU}(#1)}} 
\newcommand{\U}[1]{\ensuremath{\mathrm{U}(#1)}} 
\newcommand{\fr}[2]{\ensuremath{\frac{#1}{#2}}}
\newcommand{\x}[0]{\ensuremath{\times}}
\newcommand{\lx}[0]{\ensuremath{\ltimes}}
\newcommand{\I}{\mathrm{i}}
\DeclareMathOperator{\diag}{diag}

\begin{document}
\title{On flavor symmetries of phenomenologically viable string compactifications}

\author{Sa\'ul Ramos-S\'anchez}

\address{Instituto de F\'isica, Universidad Nacional Aut\'onoma de M\'exico,
POB 20-364, Cd.Mx. 01000, M\'exico}

\ead{ramos@fisica.unam.mx}

\begin{abstract}
Heterotic orbifolds can explain the origin of flavor symmetries 
and the flavor representations of matter fields in particle physics as a result of 
the geometric properties of the associated string states in the compact space. After a review of the method
to obtain flavor symmetries in these models, we determine the most frequent 
non-Abelian flavor symmetries appearing in promising Abelian heterotic orbifolds. Interestingly,
these symmetries correspond only to $D_4$, $\Delta(54)$ and products of these symmetries and Abelian
factors. A large set of promising models exhibits purely Abelian flavor symmetries.
We finally explore the phenomenological potential of a sample model endowed with $\Delta(54)$
assuming certain {\it ad hoc} flavon expectation values.

\end{abstract}


\section{Introduction}

One of the goals of flavor phenomenology is to discover the underlying structure in particle physics that may solve
some questions left unanswered in the SM, such as the origin of the family replication, the patterns of 
quark and lepton mixing matrices, the origin of CP violation and the absence of flavor-changing neutral 
currents (FCNC).
The field theoretic approach consists in first freely choosing a non--Abelian discrete symmetry
within $\SU3^5_{flavor}$, the largest global symmetry of the SM in the absence of Yukawa couplings,
and then introducing a number of {\it ad hoc} matter fields, some of them with unjustified expectation values (VEVs),
to fulfill different basic phenomenological constraints, such as the 
quark and lepton masses and mixings. Once these restrictions are met, this bottom-up approach delivers
a number of consequences, which frequently include interesting new physics. There are plenty of useful 
symmetries which have been thoroughly studied (see e.g.~\cite{Ishimori:2010au,King:2013eh} for a review), and it
is hard to learn which of them corresponds to the actual description of our Universe.

Looking for the origin of such symmetries might at least reduce the number of possibilities. In particular,
given the constraining environment of string theory, one may wonder what kind of flavor symmetries can emerge in
string compactifications that reproduce many properties of the standard model (SM) or its minimal supersymmetric extension (MSSM).
The first general studies of this question were~\cite{Kobayashi:2006wq,Nilles:2012cy}, in the context of orbifold
compactifications of the heterotic string perhaps due to their geometric simplicity. Around those studies, there
has been some progress in understanding the qualities of flavor symmetries arising in some phenomenologically viable
heterotic orbifolds~\cite{Kobayashi:2004ud,Ko:2007dz,Carballo-Perez:2016ooy},
their enhancements~\cite{Beye:2014nxa} and their generalizations in models endowed with magnetic fluxes~\cite{Abe:2016eyh}.
Recently, there has also been progress in the study of flavor symmetries from promising orientifold D-brane 
models~\cite{BerasaluceGonzalez:2012vb,Marchesano:2013ega}.

In this paper, we focus on the \E8\x\E8 heterotic string compactified on symmetric, toroidal \Z{N} and \Z{N}\x\Z{M} orbifolds, which are
the simplest compactifications since, among other features, the resulting space corresponds to a special point in
the Calabi-Yau moduli space, where the underlying CFT of string theory is valid. Thus, obtaining the structure 
of the couplings in the effective theory can be done by computing correlation functions of asymptotic string states
and vertex operators~\cite{Burwick:1990tu,Erler:1992gt,Choi:2007nb,Dixon:1990pc}. These computations
lead to a set of selection rules that determine which couplings among the effective fields are 
non-vanishing~\cite{Hamidi:1986vh,Dixon:1986qv,Casas:1991ac,Kobayashi:1991rp,Kobayashi:2011cw,Nilles:2013lda,Bizet:2013gf}, and that
can be expressed in terms of discrete symmetries. As we shall review in section~\ref{sec:symmetries},
based on~\cite{Kobayashi:2006wq,Nilles:2012cy}, these symmetries are the core of the flavor symmetries from heterotic orbifolds.

One question we address in this paper is what symmetries actually emerge from orbifold compactifications
that fulfill a minimal set of necessary conditions that guarantee their phenomenological viability.
Here, to be considered phenomenologically promising, an orbifold model must yield the SM gauge group, 
such that the hypercharge generator be
non-anomalous and (with normalization) compatible with grand unification, three generations 
of quarks and leptons, at least a couple of Higgs superfields, $H_u$ and $H_d$, 
and only vectorlike exotics w.r.t. the SM gauge group. These models are identified from a
set of several millions of consistent orbifold compactifications, what renders the task
very time-consuming. Fortunately, this search becomes accessible
thanks to tools such as the \texttt{orbifolder}~\cite{Nilles:2011aj}, which automatizes the computation
of matter spectra and the selection of the promising models.

The \texttt{orbifolder} allows one to perform a search of viable orbifold models by scanning randomly the
parameter space and comparing the spectra of the generated models. In section~\ref{sec:class},
we apply this technique to obtain a sample of randomly generated viable orbifold models, 
similar to those presented in~\cite[tab A.1]{Nilles:2014owa}, in order to identify
the most favored flavor symmetries. Our statistical results, that coincide with those of~\cite{Nilles:2014owa}
except for few cases (\Z7 and \Z2\x\Z4), help us remark that the most common non-Abelian symmetries
arising in phenomenologically promising models are the dihedral group $D_4$, $\Delta(54)$ and products of these symmetries
with Abelian discrete symmetries.

Orbifold compactifications do not only constrain the symmetry groups that can be used as flavor symmetries, but also the number of 
effective matter fields appearing in the models and their flavor representations.
Therefore, in contrast to field-theoretic models where flavor representations
can be chosen at convenience, in string models one is restricted to use the field representations given by the theory.
It is possible to go even beyond this statement: if all moduli fields determining the size and shape of the compactification
space are fixed, even the dynamics of possible flavon fields and hence their VEVs are fully set by the theory.
Unfortunately, despite recent progress in this direction~\cite{Parameswaran:2010ec, Dundee:2010sb}, it is
still too early to define whether moduli can be stabilized in this context. Consequently, the details of the phenomenology that can
be extracted from these constructions must rely on admissible (compatible with the small-field limit of SUGRA), 
yet {\it ad hoc}, VEVs of moduli and would-be flavon fields. This is the approach we follow in section~\ref{sec:pheno},
where we review the main phenomenological results of our previous work~\cite{Carballo-Perez:2016ooy}, based on the 
$\Delta(54)$ flavor symmetry, which is the favored symmetry in \Z3\x\Z3 orbifolds.

\section{Geometry of flavor in heterotic orbifolds}
\label{sec:symmetries}

We follow the discussion of~\cite{Kobayashi:2006wq,Nilles:2012cy}, that leads to the building blocks of
the flavor symmetries of heterotic orbifolds.
 
\subsection{Elements of heterotic orbifolds}
\label{subsec:orbifolds}

Here, we introduce the basic formalism of Abelian orbifolds, by using a \Z{N}\x\Z{M} as the standard. \Z{N}
orbifolds follow readily by ignoring all the elements related to the second symmetry.
\Z{N}\x\Z{M} heterotic orbifolds, with $N,M\in\Z{}$ are characterized by the quotient of
a six-dimensional torus $\maT^6$ divided by the joint action of two Abelian isometries of $\maT^6$.
Since \Z{N}\x\Z{M} must be an isometry of $\maT^6$,
the geometry of the torus defines which symmetries, up to deformations, can be moded out.
If the generators of \Z{N} and \Z{M}, $\vartheta$ and $\omega$ respectively, are considered
as rotations in six dimensions (i.e. ignoring roto-translations), all admissible choices of $N,M$ 
and the number of inequivalent torus geometries are displayed in table~\ref{tab:twists}.

The geometry of $\maT^6$ is encoded in its six-dimensional lattice $\Gamma$, whose basis vectors are
$\{e_1,\ldots,e_6\}$, which build, {\it in the simplest cases}, a space with metric given by the Cartan matrix
of a semi-simple Lie algebra. 
It is convenient to express this geometry in terms of the complex coordinates $z_1,z_2,z_3$,
on which the orbifold generators act as
\begin{eqnarray}
\label{eq:Oaction}
  \vartheta:&\;z_i\ \to\ z_ie^{2\pi\I v_i},&\qquad 0\leq|v_i|<1,\ i=1,2,3\,,\nonumber\\
  \omega:   &\;z_i\ \to\ z_ie^{2\pi\I w_i},&\qquad 0\leq|w_i|<1,
\end{eqnarray}
where the so-called {\it twist vectors} $\boldsymbol{v}=(v_1,v_2,v_3)$ and $\boldsymbol{w}=(w_1,w_2,w_3)$
are subject to the $\maN=1$ conditions $\pm v_1\pm v_2\pm v_3=0$ and $\pm w_1\pm w_2\pm w_3=0$.
In table~\ref{tab:twists} we list our choice of the twist vectors for all admissible \Z{N} and \Z{N}\x\Z{M} orbifolds.

{\small
\begin{table}[!t!]
\caption{\label{tab:twists} All admissible \Z{N} and \Z{N}\x\Z{M} orbifolds in six dimensions. We provide the corresponding 
         twist vectors and the number of allowed geometries (ignoring roto-translations) according to~\cite{Fischer:2012qj}.}
\begin{center}
\begin{tabular}{llc|llcc}
\br
\Z{N}     & twist vector         & \# geometries &  \Z{N}\x\Z{M}& twist vectors         &                      & \#geometries \\
\mr
\Z3       & $\fr{1}{3}(1,1,-2)$  & 1             &  \Z2\x\Z2    & $\fr{1}{2}(0,1,-1)$   & $\fr{1}{2}(1,0,-1)$  & 12     \\
\Z4       & $\fr{1}{4}(1,1,-2)$  & 3             &  \Z2\x\Z4    & $\fr{1}{2}(0,1,-1)$   & $\fr{1}{4}(1,0,-1)$  & 10     \\
\Z6-I     & $\fr{1}{6}(1,1,-2)$  & 2             &  \Z2\x\Z6-I  & $\fr{1}{2}(0,1,-1)$   & $\fr{1}{6}(1,0,-1)$  & 2    \\
\Z6-II    & $\fr{1}{6}(1,2,-3)$  & 4             &  \Z2\x\Z6-II & $\fr{1}{2}(0,1,-1)$   & $\fr{1}{6}(1,1,-2)$  & 4    \\
\Z7       & $\fr{1}{7}(1,2,-3)$  & 1             &  \Z3\x\Z3    & $\fr{1}{3}(0,1,-1)$   & $\fr{1}{3}(1,0,-1)$  & 5    \\
\Z8-I     & $\fr{1}{8}(1,2,-3)$  & 3             &  \Z3\x\Z6    & $\fr{1}{3}(0,1,-1)$   & $\fr{1}{6}(1,0,-1)$  & 2    \\
\Z8-II    & $\fr{1}{8}(1,3,-4)$  & 2             &  \Z4\x\Z4    & $\fr{1}{4}(0,1,-1)$   & $\fr{1}{4}(1,0,-1)$  & 5    \\
\Z{12}-I  & $\fr{1}{12}(1,4,-5)$ & 2             &  \Z6\x\Z6    & $\fr{1}{6}(0,1,-1)$   & $\fr{1}{6}(1,0,-1)$  & 1    \\
\Z{12}-II & $\fr{1}{12}(1,5,-6)$ & 1             &   &&&\\
\br
\end{tabular}
\end{center}
\end{table}
}

The action of the orbifold on $\maT^6$ is not free; that is, some points are left invariant or fixed (up to translations
$n_\alpha e_\alpha$, $n_\alpha\in\Z{}$, in the torus). For example, since $\vartheta$ y $\omega$ are only rotations, the origin is always a 
fixed point in the orbifold. Besides this trivial fixed point, there are non-trivial fixed points away from the origin.
The fixed points turn out to be curvature singularities of the compact space, but do not 
lead to undesirable gravitational effects in the four-dimensional uncompactified space, which is flat at first approximation.

Together with torus translations, the orbifold rotations build the {\it space group} $S=(\Z{N}\x\Z{M})\lx\Gamma$, 
whose elements are $g=(\vartheta^n\omega^m,n_\alpha e_\alpha)$, with $0\leq n< N$ , $0\leq m< M$ and $n_\alpha\in\Z{}$.
The action of the space group on the complex coordinates is given by
\begin{equation}
\label{eq:Oaction}
  S:\;z\ \to\ g z = \vartheta^n\omega^m z + n_\alpha e_\alpha\,.
\end{equation}
It is possible to associate each fixed point $z_f$ with a space group element, called {\it constructing element} $g_f$,
such that $z_f = g_f z_f$. We notice that for each choice of $(n,m)$ there are different choices of $\{n_\alpha\}$ 
leading to different fixed points. The fixed points are truly inequivalent if their corresponding constructing elements
belong to different conjugacy classes within $S$. So, at the end, for each {\it sector} $(n,m)$ there is a finite
number of fixed points.

Once the generic geometrical aspects of the compactification in six dimensions have been set, these must be embedded 
into the gauge degrees of freedom of the heterotic strings. We consider here the $\maN=1$ \E8\x\E8 heterotic string.
Modular invariance of the partition function demands that each orbifold twist be embedded either as a rotation or as  
a shift vector in the 16 dimensions of \E8\x\E8, and that the $\maT^6$ translations be connected to
a so-called Wilson line. We denote the shift vectors as $V,W$ for the embedding of $\vartheta,\omega$, and
the Wilson lines as $A_\alpha$, $\alpha=1,\ldots,6$, as the embedding of $e_\alpha$. Constructing elements in $S$
are then embedded into the gauge degrees of freedom as
\begin{equation}
\label{eq:embedding}
  g = (\vartheta^n\omega^m,n_\alpha e_\alpha) \quad \hookrightarrow \quad V_g \equiv nV+mW+n_\alpha A_\alpha\,.
\end{equation}

The gauge embedding is subject to some constraints. First, both $V$ and $W$ must be consistent
with a \Z{N}\x\Z{M} action. This amounts to requiring e.g. that $N V$ 
must lie in the root lattice of \E8\x\E8, $\Lambda$. Analogous conditions must be imposed to $W$. Secondly,
Wilson lines must be consistent with the torus geometry and the orbifold action on it.
The fact that the $\Gamma$ basis vectors $e_\alpha$ are in general related by the action of $\vartheta$ 
and $\omega$ translates to relations among different $A_\alpha$. Furthermore, from these considerations, 
just as shift vectors, Wilson lines $A_\alpha$ have an order $N_\alpha$, such that $N_\alpha A_\alpha \in \Lambda$ 
(without summation over $\alpha$).
Finally, modular invariance additionally imposes in \Z{N}\x\Z{M} heterotic orbifolds that~\cite{Ploger:2007iq}
\begin{eqnarray}
\label{eq:ModInv}
N\,(V^2-v^2) = 0\mod2\,,           &\quad& N_\alpha\,(V\cdot A_\alpha) = 0\mod 2\,,\quad\alpha=1,\ldots,6\,,\\
M\,(W^2-w^2) = 0\mod2\,,           &\quad& N_\alpha\,(W\cdot A_\alpha) = 0\mod 2\,, \nonumber\\
M\,(V\cdot W-v\cdot w) = 0\mod2\,, &\quad& N_\alpha\  A_\alpha^2 = 0\mod 2\,,\nonumber\\
                                   &\quad& \text{gcd}(N_\alpha,N_\beta)\,(A_\alpha\cdot A_\beta) = 0\mod 2\,,\quad \alpha\neq\beta\,.\nonumber
\end{eqnarray}
If all the previous requirements are fulfilled, all ingredients can be used to compactify a heterotic string.
Note that $A_\alpha=0$ $\forall\alpha$ is always a possibility.

We now turn to the matter fields $\Phi$ in orbifold compactifications. These correspond to closed string states $|\Phi\rangle$ 
that are invariant under the orbifold action and that must be massless because the mass of massive strings 
is some factor of the Planck scale, $M_{pl}$, and thus too high to appear in the effective theory at low energies.
Closed strings comprise left-movers, that equip physical states with gauge quantum numbers,
and right-movers providing the so-called $H$-momentum~\cite{Kobayashi:2011cw}, which is basically the 
momentum of the state in the compact dimensions.

In heterotic orbifolds, bulk or {\it untwisted fields} 
correspond to the orbifold-invariant states arising directly from the ten-dimensional closed strings of 
the uncompactified heterotic string, whose field limit is ten-dimensional $\maN=1$ supergravity endowed 
with an \E8\x\E8 Yang-Mills theory. Thus, the four-dimensional gauge superfields, generating
the unbroken gauge group $\maG_{4D}\subset\E8\x\E8$, and some four-dimensional matter
states with non-trivial gauge quantum numbers under $\maG_{4D}$ live in the bulk of a heterotic orbifold.

Additionally, there are the so--called {\it twisted fields}, which arise from strings that 
are closed only due to the action of the orbifold. Twisted fields are always localized at singularities of the orbifold
and are thus related to a constructing element $g$ and its embedding into the gauge degrees of freedom $V_g$ defined
in~\eqref{eq:embedding}. Since it is $V_g$ what defines the gauge quantum numbers of the localized strings,
we notice that for each sector $(n,m)$ all states have identical gauge numbers unless $A_\alpha\neq0$ for some $\alpha$,
i.e. unless there are non-trivial Wilson lines.

\subsection{Coupling selection rules}

Couplings among string states are subject to a set of constraints called {\it string 
selection rules}~\cite{Hamidi:1986vh,Dixon:1986qv,Casas:1991ac,Kobayashi:1991rp,Kobayashi:2011cw,Nilles:2013lda,Bizet:2013gf},
due to symmetries of the underlying CFT of the compactified string theory. The main
restrictions are

\ding{"0C0}\quad {\it Gauge invariance}: the sum of all gauge quantum numbers must be trivial.

\ding{"0C1}\quad {\it $H$-momentum conservation}: all $H$-momenta of the states must add up to zero.

\ding{"0C2}\quad {\it Space-group invariance}: the product of constructing elements must be trivial, which for $r$ interacting
strings means
\begin{equation}
\label{eq:spacegroupSel}
  \prod_{f=1}^r \left(\theta_{(f)},\, n^{(f)}_{\alpha} e_{\alpha}\right) \stackrel{!}{=} \Big(\mathbbm{1},\,\bigcup_f (\mathbbm{1}-\theta_{(f)})\Gamma\Big)\,,
\end{equation}
where $g_f=(\theta_{(f)},\, n^{(f)}_{\alpha} e_{\alpha})$, $\theta_{(f)}=\vartheta^{n^{(f)}}\omega^{m^{(f)}}$, corresponds to the constructing element of a twisted state
localized at $z_f$, and $\cup_f(\mathbbm{1}-\theta_{(f)})\Gamma$ is known as the {\it invariant sublattice} of fixed points.
The space group selection rule~\eqref{eq:spacegroupSel} indicates if $r$ closed strings can interact or not taking as
criterion their localizations in the compact space.

These selection rules establish for which combination of string states there is a non-zero 
correlation function, and thus a non-vanishing coupling for the associated effective fields.
We can easily see that the only selection rule that is relevant for flavor symmetries is the space-group invariance.
Notice that the rotational part of that selection rule implies
\begin{equation}
  \sum_{f=1}^r n^{(f)} = 0\mod N\,,\qquad \sum_{f=1}^r m^{(f)} = 0\mod M\,,
\end{equation}
which is a way to express invariance under a $\Z{N}\x\Z{M}$ symmetry for fields with
charges $(n^{(f)},m^{(f)})$. In explicit examples, it can be shown, as we do in the next section, that the translational part also
implies the existence of additional orbifold and geometry dependent Abelian symmetries with charges $n_\alpha^{(f)}$. 
That is, in the four-dimensional model emerging from an Abelian heterotic orbifold, space-group invariance amounts to including additional 
Abelian symmetries and assign thus appropriate discrete charges to each field in the model.

\subsection{Non-Abelian flavor symmetries}
\label{subsec:flavorsym}

\begin{figure}[!b!]
\begin{center}
\subfigure[Fixed points and charges of $S^1/\Z2$]{
\label{fig:S1overZ2}
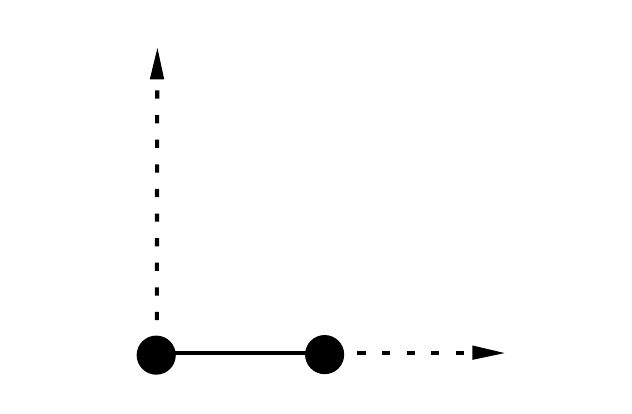
}
\hskip 1cm
\subfigure[Fixed points and charges of $\maT^2/\Z3$]{
\label{fig:T2overZ3}
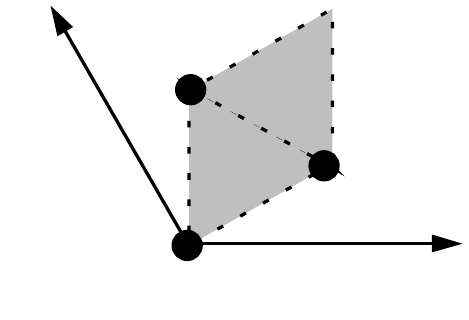
}
\caption{Geometrical origin of the $D_4$ and $\Delta(54)$ flavor symmetries in orbifolds.
         When unaffected by Wilson lines, the fixed points of an $S^1/\Z2$ ($\maT^2/\Z3$) orbifold realize an 
         $S_2$ ($S_3$) permutation symmetry. String selection rules impose an additional $\Z2\x\Z2$ ($\Z3\x\Z3$)
         symmetry based on the localization charges of twisted states. The resulting flavor symmetry is
         $S_2\lx\Z2^2=D_4$ ($S_3\ltimes\Z3^2=\Delta(54)$).}
\label{fig:D4Delta54Origin}
\end{center}
\end{figure}

As guiding examples of the origin of flavor symmetries, let us consider the cases illustrated 
in figure~\ref{fig:D4Delta54Origin}. The first figure corresponds to a \Z2 symmetry that, instead of 
being moded out of the whole $\maT^2$, divides only one compact dimension, i.e. $S^1$.
This scenario is realized in heterotic orbifolds when $\maT^6$ can be factorized as $\maT^2\x\maT^4$
and there is a Wilson line associated to one of the compact directions ($e_2$ in the depicted case).
The \Z2 acts as a reflection on the points of $S^1$, so that the elements of \Z2 that act on $z_1$ are
expressed as $\{\vartheta^0=1, \vartheta^1=-1\}$. Thus, the action of $\vartheta$ on the $\Gamma$ basis vector
reads $e_1\to -e_1$. With this at hand, we observe that two fixed points occur in the sector $\vartheta^n$
with $n=1$ and are given by the constructing elements $g_0=(\vartheta,0)$ and $g_1=(\vartheta, e_1)$, corresponding
to $z_0=0$ and $z_1=\frac12e_1$.

From the rotational part of space-group invariance, we realize that the only non-vanishing couplings are those 
satisfying $\vartheta^r=1$, i.e. when an even number of strings from the $\vartheta$ sector interact. This implies 
that states are charged with a \Z2 charge that can be either $n=0$ for untwisted sector states or
$n=1$ for states in the $\vartheta$ sector. On the other hand, the translational component
of space-group invariance requires verifying first what the invariant sublattice is. Since $(1-\vartheta)e_1=2e_1$,
then the invariant sublattice is given by any integer multiple of $2e_1$. Consider now a coupling between, say, two
twisted states, related to the constructing elements $(\vartheta,n_1^{(1)}e_1)$ and $(\vartheta,n_1^{(2)}e_1)$.
The product of these states yields $(1,(n_1^{(1)}-n_1^{(2)})e_1)$, which is equivalent to $(1,(n_1^{(1)}+n_1^{(2)})e_1)$
because they differ by a contribution of the invariant sublattice. Hence, the selection rule implies that $n_1^{(1)}+n_1^{(2)}$ 
must be even, or $\sum_f n_1^{(f)}=0\mod2$ in general,
i.e. another \Z2 that associates the charge $n_1^{(f)}$ to the twisted states. We find thus that
couplings between two twisted states are allowed only if both states lie at the same fixed point. 
We have found a \Z2\x\Z2 symmetry whose generators can be expressed in the space of the  
$\vartheta$ fixed points as the matrices $R=\diag(-1,-1)$ and $T=\diag(1,-1)$.

Finally, in the absence of Wilson lines associated with the direction $e_1$, both singularities
are indistinguishable from the point of view of a four-dimensional observer because
the gauge quantum numbers of states living at both singularities are equal. Therefore, there is a
permutation symmetry $S_2$ of the states at the singularities, whose generator can be expressed as
\begin{equation}
  P=\left(
  \begin{array}{cc}
    0 & 1\\
    1 & 0 
  \end{array}
  \right)\,.
\end{equation}
Noting that conjugating with $P$ any element of \Z2\x\Z2 generated by $R$ and $T$ yields another element
of this group, we see that \Z2\x\Z2 is a normal subgroup of the full discrete symmetry, which can thus
be written as $S_2\lx(\Z2\x\Z2)$, isomorphic to $D_4$. This is the flavor symmetry of the fields arising
from an $S^1/\Z2$ sector of a heterotic orbifold.

As we discuss in full detail in~\cite{Carballo-Perez:2016ooy}, similar considerations yield the flavor 
symmetry $\Delta(54)$ in the $\maT^2/\Z3$ orbifold, depicted in figure~\ref{fig:T2overZ3}. 
In this case, $\vartheta e_1 \to e_2$ and  $\vartheta e_2 \to -e_1-e_2$. Further, the constructing elements associated with the three fixed points of
the $\vartheta$ sector are $(\vartheta,n_1e_1)$, $n_1=0,1,2$. With the invariant sublattice 
given by the basis $\{3e_1,3e_2\}$, one can show that space-group invariance implies that only couplings of a multiple of three states
are allowed and that $\sum_f n_1^{(f)}=0\mod3$ must be satisfied.
The related Abelian symmetry is then \Z3\x\Z3, generated by $R=\diag(\rho,\rho,\rho)$ and 
$T=\diag(1,\rho, \rho^2)$, with $\rho=e^{2\pi\I/3}$. Furthermore, in the absence of Wilson lines related to 
$\maT^2$, there is a permutation symmetry $S_3$, whose generators are
\begin{equation}
  \sigma=\left(
  \begin{array}{ccc}
    0 & 0 & 1\\
    1 & 0 & 0\\
    0 & 1 & 0
  \end{array}
  \right)\,,\qquad
  \tau=\left(
  \begin{array}{ccc}
    0 & 1 & 0\\
    1 & 0 & 0\\
    0 & 0 & 1
  \end{array}
  \right)\,.
\end{equation}
The multiplicative closure of all elements reveals that the emerging flavor symmetry is $S_3\lx(\Z3\x\Z3)$,
isomorphic to $\Delta(54)$.

Another interesting aspect is that the flavor representations of matter fields are also determined by this 
structure. States associated to the fixed points of $S^1/\Z2$ can only build doublets \rep2 of $D_4$. 
In $\maT^2/\Z3$, the only matter representations in the $\vartheta$ sector
are triplets $\rep3_{11}$ of $\Delta(54)$ (in the notation of~\cite{Ishimori:2012zz}); in the $\vartheta^2=\vartheta^{-1}$ sector
the representations are conjugate, i.e. $\rep3_{12}$. Bulk states are in all cases trivial singlets of the 
corresponding flavor symmetries. No further representations appear unless states form condensates.

The permutation symmetries are broken completely if there are non-vanishing Wilson lines associated to the compact 
directions. Thus, the flavor symmetries suffer an explicit breakdown to \Z2\x\Z2 or \Z3\x\Z3 
in $S^1/\Z2$ or $\maT^2/\Z3$, respectively, when affected by non-trivial Wilson lines.  Flavor symmetries can also be broken 
spontaneously to non-Abelian subgroups if some fields
localized at the singularities develop VEVs. The details depend on the VEV structure and the number of flavon fields.

This discussion has been explicitly developed for all possible sub-orbifolds (in less than six dimensions) 
appearing in Abelian toroidal heterotic orbifolds~\cite{Kobayashi:2006wq}, resulting in a reduced number 
of family symmetries. The findings include, besides $D_4$ and $\Delta(54)$, only the symmetries
$(D_4\x D_4)/\Z2$, $(D_4\x\Z4)/\Z2$, $(D_4\x\Z8)/\Z2$ and $S_7\ltimes\Z7^6$.
As we shall shortly see, not all of these symmetries are realized in phenomenologically viable models.

\section{Favored flavor symmetries in \Z{N} and \Z{N}\x\Z{M} orbifolds}
\label{sec:class}

The tools explained in the previous section can be applied to all Abelian orbifolds. Therefore, given a
set of consistent shift vectors and Wilson lines, it is possible first to obtain the spectrum and
then to determine the flavor symmetry by inspecting which Wilson lines are trivial and what is the
orbifold action on the $\Gamma$ basis vectors associated to them.

For example, \Z3\x\Z3 orbifold models in their simplest geometry ($(1,1)$ in the nomenclature of~\cite{Fischer:2012qj}) 
admit three independent Wilson lines, each associated to one of the $z^i$ complex planes with basis vectors $e_{2i-1}$ and $e_{2i}$.
The global flavor symmetry in the absence of Wilson lines is $(S_3\x S_3\x S_3)\ltimes\Z3^5$, which is isomorphic to 
$\Delta(54)^3/\Z3$. Now, if one of the Wilson lines is non-zero, then the flavor symmetry is $(S_3\x S_3)\lx\Z3^5$ due
to the explicit breaking of one $S_3$. Two non-vanishing Wilson lines yield $S_3\lx\Z3^5$, which can be rewritten as
$\Delta(54)\x\Z3^3$. (This symmetry shall play an important role in the next section.)
And three non-trivial Wilson lines lead to a purely Abelian $\Z3^5$ flavor symmetry.

There are other more complicated cases, such as \Z6-II orbifolds, whose structure leads to the possibility of three
non-trivial Wilson lines: one related to the $\maT^2$ of the $z^2$ plane, and two associated to the $e_5$ and $e_6$ directions. 
From the orbifold action determined by the twist $v=(1,2,-3)/6$, we see that the orbifold corresponds to the factorization
$\maT^2/\Z6\x\maT^2/\Z3\x S^1/\Z2\x S^1/\Z2$, where the Wilson lines can affect the last three factors. In this case,
it is not enough to know the number of non-zero Wilson lines of a given model to figure out the flavor symmetry. One needs also
the direction with which it is associated. If a model is furnished with one non-trivial Wilson line, then the original
flavor symmetry $(S_3\x S_2 \x S_2)\lx (\Z3^2\x\Z2^3)$ is broken down to one of two possibilities,
$\Delta(54)\x D_4\x\Z2$ or $(S_2\x S_2)\lx(\Z3^2\x\Z2^3)$, isomorphic to $(D_4\x D_4)/\Z2\x\Z3^2$.
Two non-vanishing Wilson lines render the flavor symmetries $\Delta(54)\x\Z2^3$ or $D_4\x\Z3\x\Z3\x\Z2$.
And three non-zero Wilson lines lead to a $\Z3^2\x\Z2^3$ flavor symmetry.

{\small
\begin{table}[!t!]
\caption{\label{tab:T1} Statistically preferred flavor symmetries in phenomenologically viable \Z{N} and \Z{N}\x\Z{M} heterotic
         orbifolds. The first two columns of each table correspond to the label of Abelian orbifolds, following the notation of~\cite{Fischer:2012qj}.
         The third column displays the most common flavor symmetry appearing in promising models; the symmetries in
         squared brackets are non-Abelian flavor symmetries that appear less frequently in these models. Additional Abelian 
         factors are not displayed. The last column counts the number of independent non-trivial Wilson lines needed to build most frequent 
         promising models over the maximal number of Wilson lines allowed by the geometry. }
\begin{minipage}[t]{0.49\textwidth}
\begin{tabular}{llcccllcc}
\br
\Z{N}    &          & flavor symmetry     & \#  WL \\
\mr
\Z3      & $(1,1)$  & no viable model     & --     \\
\hline
\Z4      & $(1,1)$  & no viable model     & --     \\
         & $(2,1)$  & \Z4                 & 3/3    \\
         & $(3,1)$  & \Z4                 & 2/2    \\
\hline
\Z6-I    & $(1,1)$  & \Z6                 & 1/1    \\
         & $(2,1)$  & \Z6                 & 1/1    \\
\hline
\Z6-II   & $(1,1)$  & $D_4$               & 2/3    \\
         & $(2,1)$  & $D_4$               & 2/3    \\
         & $(3,1)$  & $D_4$               & 2/3    \\
         & $(4,1)$  & $D_4$               & 1/2    \\
\hline
\Z7      & $(1,1)$  & \Z7                 & 1/1    \\
\hline
\Z8-I    & $(1,1)$  & \Z8 [{\small$(D_4\x\Z8)/\Z2$}]   & 2/2    \\
         & $(2,1)$  & \Z8 [{\small$(D_4\x\Z8)/\Z2$}]   & 2/2    \\
         & $(3,1)$  & \Z8                              & 1/1    \\
\hline
\Z8-II   & $(1,1)$  & $D_4$ [{\small$(D_4\x\Z8)/\Z2$}] & 2/3    \\
         & $(2,1)$  & $D_4$                            & 1/2    \\         
\hline
\Z{12}-I & $(1,1)$  & \Z{12}              & 1/1    \\
         & $(2,1)$  & \Z{12}              & 1/1    \\
\hline
\Z{12}-II& $(1,1)$  & \Z{12} [{\small$D_4$}]           & 2/2    \\
\br
\end{tabular}
\end{minipage}
\begin{minipage}[!t!]{0.49\textwidth}
\begin{tabular}{llcccllcc}
\br
\Z{N}\x\Z{M}                    &          & flavor symmetry                    & \#  WL \\
\mr
\Z2\x\Z2                        & $(1,1)$  & $(D_4\x D_4)/\Z2$ [{\small $D_4$}] & 4/6  \\
\hline
\Z2\x\Z4                        & $(1,1)$  & $D_4\x(D_4\x\Z4)/\Z2$              & 2/4   \\
                                &          & [$(D_4\x D_4)/\Z2$]                & 2/4   \\
                                &          & [$D_4$ or $(D_4\x\Z4)/\Z2$]   & 3/4   \\
\hline
\Z2\x\Z6-I                      & $(1,1)$  & \Z2\x\Z6 [{\small $D_4$}]        & 1/2   \\
\hline
\Z2\x\Z6-II                     & $(1,1)$  & no viable model                  & --    \\
\hline
\Z3\x\Z3\phantom{$1^{1^1}$}     & $(1,1)$  & $\Delta(54)$                     & 2/3   \\
                                &          & [$(\Delta(54)\x\Delta(54))/\Z2$] & 1/3   \\
\hline
\Z3\x\Z6                        & $(1,1)$  & \Z3\x\Z6 [{\small$\Delta(54)$}]          & 1/2   \\
\hline
\Z4\x\Z4                        & $(1,1)$  & $((D_4\x\Z4)/\Z2)^2/\Z2$         & 1/3   \\
                                &          & [$(D_4\x\Z4)/\Z2$]               & 2/3   \\
\hline
\Z6\x\Z6                        & $(1,1)$  & \Z6\x\Z6                         & --    \\
\br
\end{tabular}
\vskip 31mm
\phantom.
\end{minipage}
\end{table}
}

This exercise can be performed for phenomenologically promising heterotic orbifolds. The use of the 
\texttt{orbifolder}~\cite{Nilles:2011aj} leads to large sets of models with {\it a priori}
defined properties. Particularly, one can obtain models with the following features:
1) SM gauge group with non-anomalous and correctly normalized hypercharge,
2) three SM generations,
3) at least a couple of Higgses $H_u$ and $H_d$, and
4) no chiral exotics.
These models will be considered phenomenologically viable. Then, identifying their flavor symmetries can help
envisage the phenomenological potential of these constructions.

We have performed non-exhaustive scans of all orbifolds listed in tables~\ref{tab:T1}, with the geometries
specified in the second column of each table in the notation of~\cite{Fischer:2012qj}. That is, we have
used all possible geometries (ignoring roto-translations) of \Z{N} orbifolds and only the simplest ones
for \Z{N}\x\Z{M} orbifolds. Our results, that shall further detailed elsewhere, are statistically equivalent to
those of~\cite[tab A.1]{Nilles:2014owa}, except in the \Z7 orbifold where we do find one promising model with 
one Wilson line and in the \Z2\x\Z4 orbifold, where we find that most of the models are endowed with two 
non-vanishing Wilson lines, instead of three.

Out of nearly 8,000 models, about 30\% of them arise from \Z2\x\Z4 models and around $10+10$\% from \Z4\x\Z4 
and \Z6\x\Z6 orbifolds. About 35\% of the models saturate the number of non-vanishing Wilson lines. In these cases,
the flavor symmetries are purely Abelian, rendering the models less promising. These appear mostly in \Z{12}-I
and \Z6\x\Z6 orbifolds, which may suggest that some efforts in the literature about \Z{12}-I orbifolds should be redirected.

We have examined the most common flavor symmetries in each orbifold geometry. Our results are listed in 
table~\ref{tab:T1}. There, we display only the non-Abelian flavor symmetries identified, omitting smaller Abelian
factors, for the sake of simplicity. In the cases where only Abelian flavor symmetries appear, we present only the 
largest Abelian group. The symmetries in squared brackets correspond to less favored non-Abelian flavor symmetries that
arise in our models.

In many cases where most of the models exhibit Abelian flavor symmetries, there are also some viable
models with non-Abelian flavor symmetries. For example, in \Z8-I with torus geometry $(2,1)$, besides the 
flavor group \Z8, there are about 25\% of the models endowed with a $(D_4\x\Z8)/\Z2$ flavor symmetry.

Globally, we also find that about 30\% of the models yield a $D_4$ symmetry multiplied by purely Abelian factors.
Around 21\% of the models have non-trivial combinations of $D_4$ with itself and other Abelian factors or quotients, 
and close to 5\% of the models are furnished with a $\Delta(54)$ flavor symmetry. That is, considering the models that 
exhibit these flavor symmetries together with those endowed with purely Abelian flavor groups, we obtain around 90\% of
all our models.

\section{Phenomenological consequences of flavor symmetries in string models}
\label{sec:pheno}

Our results of the previous section show that most promising heterotic orbifolds enjoy a $D_4$ non-Abelian
flavor symmetry. The phenomenological potential of this symmetry in this context has been studied in detail 
in~\cite{Ko:2007dz,Lebedev:2007hv}, where admissible Yukawa textures for quarks and (charged and neutral) leptons
as well as promising supersymmetric soft terms (that help avoid FCNC in supersymmetric models) are achieved.
It is interesting that the same flavor symmetry appears frequently in appealing D-brane models~\cite{Marchesano:2013ega}.

We have also found that $\Delta(54)$ is the second most favored non-Abelian flavor symmetry appearing in viable heterotic orbifolds
and it has been largely ignored even from a bottom-up approach. Thus, it is necessary to study its phenomenological 
consequences. To do so, we recover here the main properties of the \Z3\x\Z3 orbifold model
studied in~\cite{Carballo-Perez:2016ooy}. In that work, one of the about 800 promising models was chosen due to its 
simplicity. The model is defined by the shift vectors
\begin{subequations}
\label{eqs:shifts_model9}
\begin{eqnarray}
  3 V &=& \left(-\tfrac{1}{2}, -\tfrac{1}{2}, -\tfrac{1}{2}, -\tfrac{1}{2}, \tfrac{1}{2}, \tfrac{1}{2}, \tfrac{1}{2}, \tfrac{1}{2}; -2, 0, 0, 1, 1, 1, 1, 4\right)\,, \\
  3 W &=& \left( 0, 1, 1, 4, 0, 0, 1, 1; 1, -1, 4, -4, -1, 0, 0, 1\right)\,,
\end{eqnarray}
\end{subequations}
and the Wilson lines
\begin{subequations}
\label{eqs:WL_model9}
\begin{eqnarray}
  3 A_{1} = 3 A_2 & = & \left(-\tfrac{7}{2}, -\tfrac{3}{2}, \tfrac{9}{2}, \tfrac{7}{2}, -\tfrac{7}{2}, -\tfrac{3}{2}, \tfrac{5}{2}, \tfrac{7}{2};  -3, 0, -2, 0, -2, -4, 3, -2\right)\,,\\
  3 A_{3} = 3 A_4 & = & \left( 3, 3, -3, -2, -1, 2, 4, -4;  -3, 1, -1, -4, 1, 1, 4, 1\right)\,.
\end{eqnarray}
\end{subequations}
These parameters yield the unbroken gauge group $\SU3_C\x\SU2_L\x\U1_Y\x[\SU2\x\U1^{11}]$, where the
additional \SU2 factor is considered hidden because no SM--field carries a charge under that group.
All fields in the spectrum are charged under the additional \U1 factors, but these shall be broken by flavon VEVs.
The flavor symmetry is $\Delta(54)\x\Z3^3$.

The relevant matter spectrum is displayed in table~\ref{tab:matterfields}, where the first eight columns refer to 
SM matter states, including the Higgs fields and right-handed neutrinos, and the last five columns display the properties of flavon fields.
In this model, SM fermion fields transform as triplets of the $\Delta(54)$ flavor symmetry 
while the Higgs fields do not transform because they are bulk fields.

\begin{table}[!b!]
\begin{center}
\begin{tabular}{|c|c|c|c|c|c|c||c|c||c|c|c||c|c|}
\hline
$\phantom{A^{A^{A^A}}}$& $Q_i$ & $\bar{d}_i^c$ & $\bar{u}_i^c$ & $L_i$ & $\bar{e}_i^c$ &$\bar{\nu}_i$& $H_u$ & $H_d$ & $\phi^u_i$ & $\phi^{(d,e)}_i$ & $\bar{\phi}^\nu_i$ & $s^u$ & $s^{(d,e)}$\\
\hline
$\Delta(54)$ &$\rep3_{11}$ & $\rep3_{11}$  & $\rep3_{11}$  & $\rep3_{11}$  & $\rep3_{11}$&  $\rep3_{12}$  & $\rep1_0$ & $\rep1_0$  & $\rep3_{11}$ & $\rep3_{11}$ & $\rep3_{12}$ & $\rep1_0$   & $\rep1_0$ \\
\hline
$\Z3^{(1)}$  & $\rho$     & $1$          & $\rho$       & $1$          & $\rho$   &  $1$          & $1$      & $1$       & $1$         & $1$         & $1$         & $\rho$   & $\rho^2$ \\
\hline
$\Z3^{(2)}$  & $1$        & $\rho^2$     & $1$          & $\rho^2$     & $1$        &  $\rho$     & $1$      & $1$       & $1$         & $\rho$    & $\rho$    & $1$        & $1$ \\
\hline
$\Z3^{(3)}$  & $\rho$     & $1$          & $\rho$       & $1$          & $\rho$   &  $1$          & $1$      & $1$       & $1$         & $1$         & $1$         & $\rho$   & $\rho^2$ \\
\hline
\end{tabular}
\caption{Flavor representations for the SM matter and flavon fields in a \Z3\x\Z3 sample model. 
         The \Z3 charges are defined by the field localizations with $\rho=e^{2\pi\I/3}$.}
\label{tab:matterfields}
\end{center}
\end{table}

Since the SM matter fields are charged under flavor symmetries, only the presence of the properly charged
$s$ and $\phi$ flavon fields allows for Yukawa couplings in the (non--renormalizable) superpotential, 
given by
\begin{eqnarray}
\label{eq:Wyuk}
W_{Y} &=& y_{ijk}^u Q_i H_u \bar{u}_j \phi^u_k s_u + y_{ijk}^d Q_i H_d \bar{d}_j \phi^{(d,e)}_k s^{(d,e)} + y_{ijk}^e L_i H_d \bar{e}_j \phi^{(d,e)}_k s^{(d,e)} \\
      &+& y_{ijkl}^\nu L_i H_u \bar{\nu}_j+ \lambda_{ijk} \bar\nu_i \bar\nu_j \bar\phi^\nu_k\,, \qquad\qquad\qquad\qquad i,j,k=1,2,3,  \nonumber
\end{eqnarray}
where the summation over repeated indices must follow the rules of the product of $\Delta(54)$ representations
that lead to invariant singlets. We see from this superpotential, that
quarks and charged leptons acquire masses through dimension--6 operators, 
and the Dirac neutrino masses as well as the right-handed Majorana neutrino masses are generated 
at renormalizable level.  This is optimal because guarantees that the largest masses are those of right-handed neutrinos.

Assuming that most properties are conserved even after supersymmetry breakdown, we observe that choosing some 
{\it ad hoc} hierarchical flavon--VEV alignments results
in the following flavor phenomenology features:
\begin{itemize}
 \item correct masses for quarks and charged leptons;
 \item proper Gatto-Sartori-Tonin relation in the quark sector (although the other two mixing angles are very small);
 \item a mass relation between the down--quark sector and the charged leptonic sector
       \begin{equation}
       \label{eq:massrelation}
          \frac{m_s-m_d}{m_b} \ \stackrel{!}{=} \ \frac{m_\mu-m_e}{m_\tau};
       \end{equation} 
 \item compatibility (only) with normal hierarchy of neutrino masses;
 \item smallest neutrino mass of order $6-7$ meV; 
 \item total neutrino masses of order $65-70$ meV; and
 \item PMNS matrix compatible with current constraints (atmospheric and reactor 
       mixing angles are in the $3\sigma$ region of the global best fit), 
       with the atmospheric mixing angle greater than $45$ degrees.
\end{itemize}
Interestingly, an inverted hierarchy being disfavored as well as the atmospheric 
mixing angle lying in the second octant, are features compatible with recent
findings of the T2K collaboration~\cite{Abe:2017uxa}. These results render the neutrino sector of $\Delta(54)$ heterotic orbifolds
much more promising than the other sectors and let us assert that \Z3\x\Z3 heterotic orbifolds and $\Delta(54)$ as a flavor 
symmetry provide a fertile playground for useful phenomenology.

\section{Concluding remarks}

In this paper, we firstly reviewed the geometric origin of flavor symmetries. They appear as a result of the 
effective properties of matter fields due to their localization in the compact space. The selection 
rules that define the non-zero couplings among fields are the key string ingredient in the construction of
flavor symmetries. Non-Abelian symmetries turn out to be semi-direct products of a permutation symmetry, 
arising from the degeneracy of localized states in the absence of non-trivial Wilson lines, multiplied with 
some Abelian discrete symmetries inherited from the selection rules.

We have then performed a large scan of promising heterotic orbifolds and identified the statistically favored 
flavor symmetries in those models. We find that in about 50\% of the explored models the $D_4$ 
flavor symmetry is preferred. The second most frequent non-Abelian flavor symmetry in phenomenologically
viable models is $\Delta(54)$. It is somewhat surprising that about 35\% of the models carry purely 
Abelian groups as flavor symmetries.

Since $D_4$ has been explored in the past, here we reviewed the main features of a \Z3\x\Z3
orbifold model giving rise to the SM particle spectrum that forms $\Delta(54)$ flavor representations. 
Exploiting the couplings structure of the model and choosing some {\it ad hoc} flavon VEVs, 
the model reproduces all particle masses, but exhibits issues to yield the 
right quark and charged-lepton mixing angles. Yet the neutrino sector, that allows for see-saw neutrino masses,
can meet the experimental constraints for all mixing angles, favors a normal hierarchy in the neutrino sector, 
and provides promising values for the left-handed neutrino masses. One is thus encouraged to explore other models with 
this symmetry to exhaust the phenomenological potential of $\Delta(54)$.

Two reasonable questions arise from this study. Given the most favored symmetries, can we provide a generic
recipe for the flavor phenomenology arising from string models? Is it possible to falsify some of these 
promising heterotic orbifolds on the basis of their flavor symmetries? In clarifying these general questions,
it would also be convenient to perform a more exhaustive scan, including all possible toroidal geometries of \Z{N}\x\Z{M}
orbifolds, roto-translations, and the enhancement(s) of these symmetries at special modular values. 
These open challenges shall be the aim of future works.

\section*{Acknowledgments}
I thank B. Carballo-Pérez, E. Peinado and Y. Olguín-Trejo for fruitful collaborations on this subject.
This work was partly supported by DGAPA-PAPIIT grant IN100217 and CONACyT grant F-252167.
I would like to thank the ICTP for the kind hospitality and support received
through its Junior Associateship Scheme during the realization of this work.

\section*{References}

\providecommand{\newblock}{}

\end{document}

%% file: Z2fixedpoints.pdf_t
\begin{picture}(0,0)%
\includegraphics{Z2fixedpoints.pdf}%
\end{picture}%
\setlength{\unitlength}{10359sp}%
\begingroup\makeatletter\ifx\SetFigFont\undefined%
\gdef\SetFigFont#1#2#3#4#5{%
  \reset@font\fontsize{#1}{#2pt}%
  \fontfamily{#3}\fontseries{#4}\fontshape{#5}%
  \selectfont}%
\fi\endgroup%
\begin{picture}(1154,750)(1759,114)
\put(2339,278){\makebox(0,0)[lb]{\smash{{\SetFigFont{10}{12.0}{\rmdefault}{\mddefault}{\updefault}$n_1=1$}}}}
\put(1826,276){\makebox(0,0)[lb]{\smash{{\SetFigFont{10}{12.0}{\rmdefault}{\mddefault}{\updefault}$n_1=0$}}}}
\put(2690,174){\makebox(0,0)[lb]{\smash{{\SetFigFont{10}{12.0}{\rmdefault}{\mddefault}{\updefault}$e_1$}}}}
\put(1944,729){\makebox(0,0)[lb]{\smash{{\SetFigFont{10}{12.0}{\rmdefault}{\mddefault}{\updefault}$e_2$}}}}
\end{picture}%

%% file: Z3fixedpoints.pdf_t
\begin{picture}(0,0)%
\includegraphics{Z3fixedpoints.pdf}%
\end{picture}%
\setlength{\unitlength}{11188sp}%
\begingroup\makeatletter\ifx\SetFigFont\undefined%
\gdef\SetFigFont#1#2#3#4#5{%
  \reset@font\fontsize{#1}{#2pt}%
  \fontfamily{#3}\fontseries{#4}\fontshape{#5}%
  \selectfont}%
\fi\endgroup%
\begin{picture}(1070,730)(1617,51)
\put(1944,115){\makebox(0,0)[lb]{\smash{{\SetFigFont{11}{13.2}{\rmdefault}{\mddefault}{\updefault}$n_1=0$}}}}
\put(2425,130){\makebox(0,0)[lb]{\smash{{\SetFigFont{11}{13.2}{\rmdefault}{\mddefault}{\updefault}$e_1$}}}}
\put(2204,277){\makebox(0,0)[lb]{\smash{{\SetFigFont{11}{13.2}{\rmdefault}{\mddefault}{\updefault}$n_1=1$}}}}
\put(1632,580){\makebox(0,0)[lb]{\smash{{\SetFigFont{11}{13.2}{\rmdefault}{\mddefault}{\updefault}$e_2$}}}}
\put(1833,489){\makebox(0,0)[lb]{\smash{{\SetFigFont{11}{13.2}{\rmdefault}{\mddefault}{\updefault}$n_1=2$}}}}
\end{picture}%